\documentstyle[11pt,mymoriond,epsfig]{article}

\def\be{\begin{equation}}
\def\ee{\end{equation}}
\def\bea{\begin{eqnarray}}
\def\eea{\end{eqnarray}}

\begin{document}
\vspace*{4cm}
\title{Higgs: KeV precision and CP violation}

\author{ W.J. Murray }

\address{Rutherford Appleton Laboratory, Chilton, Didcot, Oxon. OX11 0QX, UK}

\maketitle
\abstracts{
The prospects for a muon collider operated as a Higgs factory are reviewed.
The large muon mass means that the s-channel Higgs production mechanism
is available, and simultaneously suppresses bremsstrahlung so that the
beam energy spread can be kept to the MeV level required to exploit
this. Thus this is the only machine  which  can make a direct scan
over the Higgs resonance, and make an extraordinary mass measurement.
Further possibilities such as a scan of the H and A of supersymmetry and
CP violation are also mentioned.
}

\section{Reminder of muon colliders}

The muon collider\cite{ref:status_report,cern99} is the only way of
pushing the energy frontier
beyond the region of applicability of electron colliders, while retaining
the advantages of a point-like projectile. The mass is 200 times that of an
electron, so storage rings  behave like those of proton machines
while
a beam energy spread as low at $10^{-5}$ may be possible, which means that
narrow resonances can be scanned. The lack of beamstrahlung means
that thresholds are  clean, and the energy can be  measured to $10^{-6}$ via
g-2.

This measurement of the Higgs mass benefits  from the coupling 
$g_{h\mu\mu}$, which gives a cross-section for s-channel production
40,000 times the electron equivalent.
This allows a direct scan, giving the mass and width. This possibility 
depends upon a Higgs below the W threshold, so it is encouraging that EW
fits\cite{ref:SM-fits} and direct observation\cite{ref:heroes} appear to
favour this. Indeed one study\cite{cern99} found 115~GeV as the optimal
Higgs mass for a muon collider.

The disadvantage of a muon collider is of course the muon lifetime of
2.2~$\mu$s, which means that muon production and cooling has to be 
performed on a similar time-scale. Furthermore the electrons from the
muon decay will constitute a serious detector background. Both these
difficulties mean that we wish to maximize the luminosity per amp
of muon current, which implies cooling the beams as much as possible.

The
decay of a high energy muon beam provides an ideal source of neutrinos
for the study of the neutrino mixing. Such a project is 
simpler than a muon collider, particularly with regard to cooling, 
and it will therefore probably be built first.  Its construction will
bring advances in the techniques required for the collider, and it will
both  serve as an important proof of principle and probably be directly
used in the collider construction.

\section{Overview of the accelerator components}

\begin{figure}[h]
\center
\vspace*{-1cm}
\epsfig{file=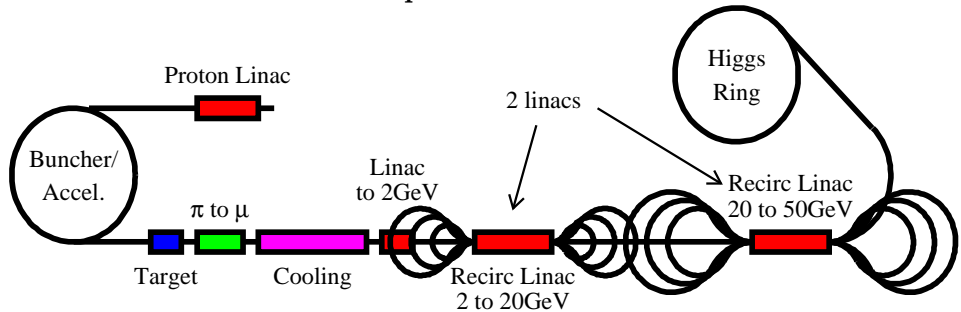,width=0.8\textwidth}
\caption{The conceptual layout of a muon collider complex.
\label{fig:scheme}}
\end{figure}

Figure~\ref{fig:scheme} shows the essential components of a muon collider.
 A proton source
provides power in the megawatt range on a target, which is optimized
for pion production. The pions are allowed to decay to muons, which
are then rapidly cooled with ionization cooling. They are then accelerated
to the required energy, probably in a recirculating linear accelerator
and fed into a collider ring. The whole procedure will repeated many
times per second.

\subsection{Proton driver}
\label{subsec:factory}

The proton source must provide a large amount of power, of the order of 
four megawatts at an energy large enough to produce pions copiously. 
The exact energy of the beam is under discussion,
with possibilities ranging from 2.2~GeV to 24~GeV, and the
HARP experiment\cite{ref:harp} at CERN is measuring pion production and
will be important in making this choice.
 The pulse length must be order 1~ns or so, in order to reduce the
phase space of the outgoing muons, but there is no advantage in
reducing it below this, because the time jitter introduced by the
$\pi$ to $\mu$ decay is of the same order.

The proton driver can either be a linac or a rapid cycling
synchrotron.
The CERN proposal for a Superconducting Proton Linac\cite{ref:spl} is the best example
of the former. This accelerator would re-use the LEP superconducting
cavities, supplemented by new cavities designed to work with lower velocity
particles. It could in principle deliver even more than 4~MW, but the
energy is limited to 2.2GeV. Furthermore it needs bunching and compression
if the small time structure is to be achieved. 

A synchrotron solution would be able to 
operate at higher beam energy, and this has the advantage that less protons
are required, and the phase space density requirements are easier. Thus
it may be more suitable for delivering a few high intensity muon pulses which
are required for a collider, rather than many small ones which might
be satisfactory for a muon neutrino source.

\subsection{Target and pion collection}

To optimize the muon rate, it is important to collect as large a fraction
of the pions produced as possible. The peak kinetic energy is rather
low, or order of the pion mass, independent of the beam power, because
most pions are produced through secondary interactions. Such pions
cannot penetrate much material. However, it is important to use a large
fraction of the energy in the proton beam, and this requires a thick 
target. The solution is to use a rod-shaped target, so that pions with
any significant $P_T$ are emitted from the sides of the target. A radius
around 5mm seems to be optimal for a fairly dense target.

\begin{figure}
\center
\epsfig{file=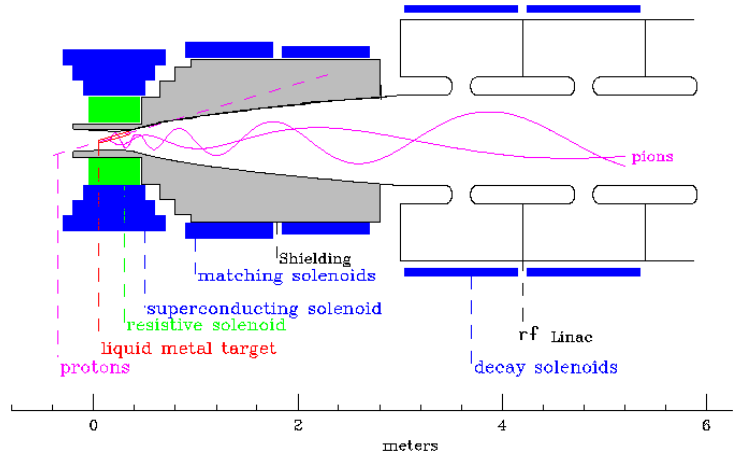,width=0.6\textwidth}
\caption{Cartoon of the pion target design. See text for details.
\label{fig:target}}
\end{figure}

The target is either inside a solenoidal magnetic field or in front
of a magnetic focusing horn, which serve to confine the pions
into a drift volume. If a solenoid is used, as shown in 
figure~\ref{fig:target} the field will be of
order 20 Tesla, which encloses pions with a transverse momentum below
225~MeV within a radius of 8~cm.

The target must be able to cope with the proton power of perhaps 4~MW.
In such a small target this gives a very large heating, and while
solid solutions\cite{ref:bennett} are still under active investigation,
the preferred design is a liquid metal jet. 
The jet is of course a conductor, and it remains to be shown that it can be
injected into such a strong magnetic field without disruption.
The jet will certainly be disrupted by the beam, but reforms 
 in around 20~ms, ready for the next proton pulse.
The proposed system has a difficult combination of magnetic fields, heat
transfer, mechanical stress and radiation exposure.
%, and presents one
%of the most challenging areas of the collider project. 

\subsection{Decay and phase rotation}

The pions must now be allowed to decay to muons which are to be cooled
and accelerated. These later steps require that the momentum spread of
the beam be reduced so that the bunch does not diverge. The simplest 
way of doing this in principle is to drift for some tens of meters, during
which the decay occurs and a correlation between velocity and arrival 
time is created. A matched phase-rotating RF system can then be employed
to decelerate the first arriving fast muons and accelerate the slowest ones.
That then gives a long bunch with relatively uniform energy.

A more sophisticated solution is to chop the bunch into several sub-bunches
which are differentially accelerated, as above, but also given different
paths so that they all arrive at the next stage at the same time. The
combination of the sub-bunches is by no means easy, and it will inevitably
increase the transverse beam emittance, but this is relatively easy to
reduce afterwards with ionization cooling.

\subsection{Cooling}

The rapid muon decay means that traditional cooling techniques are too slow,
but fortunately ionization cooling seems to provide an answer. The principle is
that muons loose energy when they pass through matter, and they can be 
re-accelerated in the longitudinal direction. This provides a net reduction
in the transverse momentum, although not longitudinally.

\begin{equation}
\frac{d \epsilon_n}{dx} = - \frac{1}{\beta^2} \frac{dE_\mu}{dx}
 \frac{\epsilon_n}{E_\mu}
        + \frac{\beta_\perp (0.014)^2}{2 \beta^3 E_\mu m_\mu L_R} \\
\label{eq:cool}
\end{equation} 

Where $\epsilon_n$ is the emittance in one transverse direction
$\sigma_x \sigma_{P_x}/m_\mu c$,  $\beta_\perp$ is the 
betatron function of the absorber and $L_R$ is the radiation length.
The first term corresponds to ionization cooling, and the second to warming
due to multiple scattering.
There is an equilibrium when 
\begin{equation}
\epsilon_n = \frac{\beta_\perp}{\beta L_R(dE_\mu/dx)}
\label{eq:equilibrium}
\end{equation} 

The optimal cooling therefore calls for minimum $\beta_\perp$. 
This is clear: the minimum
in the beta function corresponds to a maximum in the beam divergence, and
at this point the contribution from multiple scattering is least important.
However, we also wish a material which maximizes the product of 
radiation length and $dE_\mu/dx$, which will be achieved for low Z.
This product (evaluated at the minimum
of the $dE/dx$ curve), is 253
for Hydrogen, while for lithium, which is easier to handle and a conductor,
it is 131. There is thus almost a factor of two advantage in using hydrogen,
and heavier elements  are correspondingly worse. 
Another possibility is LiH, which has $L_R(dE_\mu/dx)$ of 137,
and could be a useful for an  absorber with a complex shape.

For the first stages of cooling the beam is large, and the ultimate
limit is not important, and hydrogen is a good choice of material. For the
final cooling stages it may be that higher fields can be created by the
use of a lithium lens, with kAmps of current flowing through it, and
that a low $\beta_\perp$ be more  advantageous then using hydrogen.

\begin{figure}[ht]
\center
\unitlength1cm
\begin{picture}(15.5,7.5)
\put(0,0){\epsfig{file=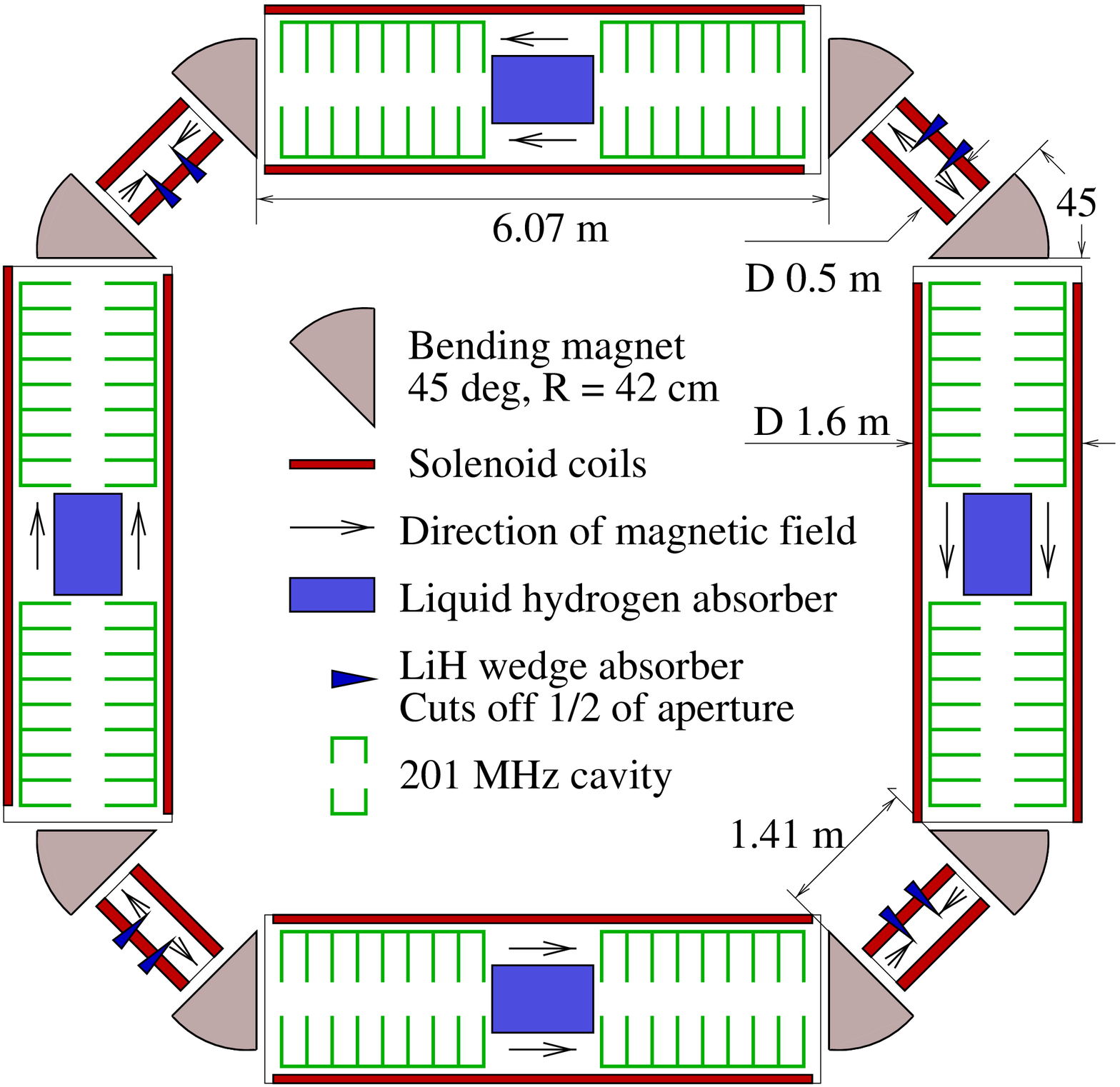,width=0.48\textwidth}}
\put(7.75,1){\epsfig{file=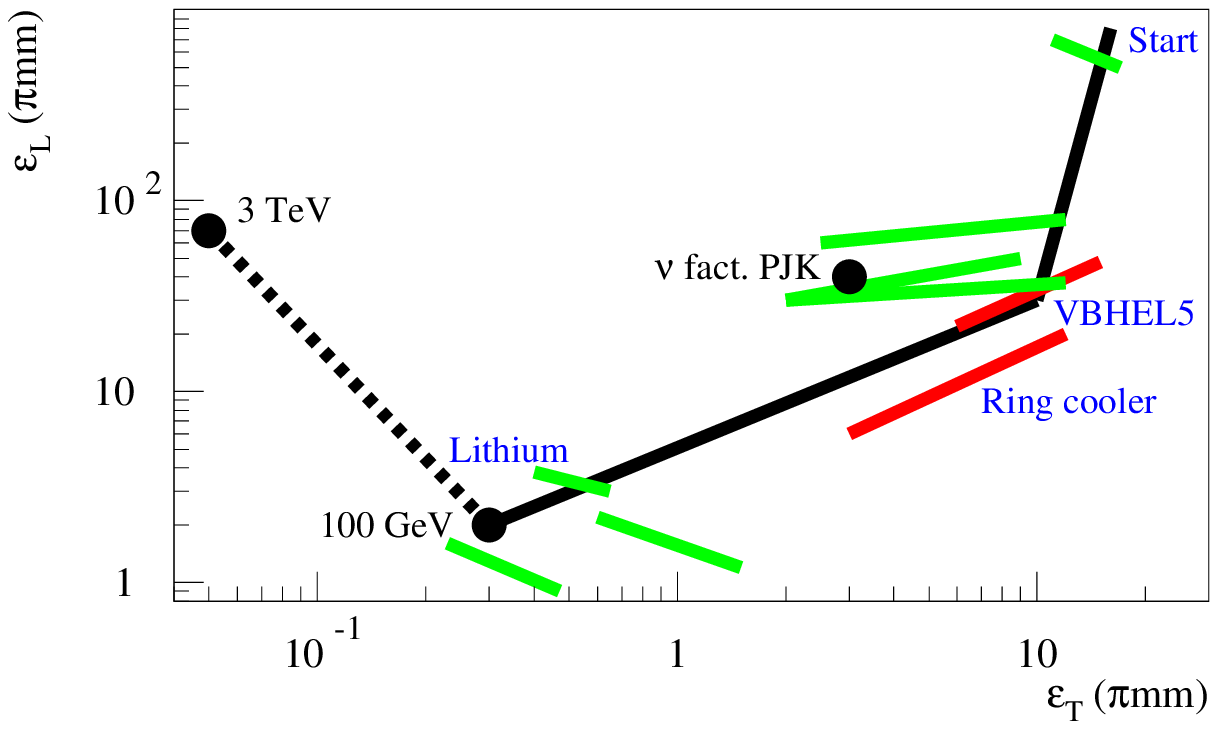,width=0.50\textwidth}}
\put(0.2,6.5){\mbox{\bf a)}}
\put(7.9,6.5){\mbox{\bf b)}}
\end{picture}
\caption{a) Sketch of a ring cooler, designed by V. Balbekov. Note the
alternate field flips and shaped LiH absorbers. 
b) shows the cooling required for the muons from production at
pion decay to use in a collider. The x and y axes are the transverse and
longitudinal emittance, and  the grey lines show the effect of various
designs of cooling channel.}
\label{fig:cooling}
\end{figure}

Cooling is required in both transverse and longitudinal directions, 
but the latter cannot  be achieved by ionization cooling directly. 
Instead it is necessary to exchange emittance between transverse and 
longitudinal components, whilest cooling transverse. This is usually
thought to be done using a magnetic field to differentially deflect
them beam, and then an absorber whose thickness varies with position
so that a greater thickness is seen by the higher energy particles.
However, detailed designs are difficult, as scatterings and imperfections
tend to warm the beam.

A promising
recent development is the ring coolers\cite{ref:balbekov}, an example
of which is shown in figure~\ref{fig:cooling}~a). These seem to be 
provide genuine six dimensional cooling despite allowing for windows
and tails of scattering. However, a ring needs fast kickers to 
inject and eject the beam, and at present there is no space available
for these.

A summary of cooling\cite{ref:fernow-2001} desired and designed can be seen in
figure~\ref{fig:cooling}~b). There are many gaps in the chain, but the 
ring coolers and lithium lens devices do seem to be pieces of the
overall cooling scheme.

\subsection{Acceleration and collider}

The acceleration is not in principle a great difficulty. To get to say 57.5~GeV
per beam for a Higgs factory will require linear accelerators, but these 
can be recycling, either in a racetrack or dog-bone design. The latter seems
to offer the best price for a specified performance, as the same accelerating
cavities are used by the muons in both directions, meaning that half
as much RF is required. 

%\subsection{Collider ring}

\begin{table}[ht]
\caption{Possible parameter sets for the collider. Higher energy machines are
shown for comparison.\label{ta:collider}}
\begin{center}
\begin{tabular}{|l|c|c|cc|} 
  \hline
CoM energy       & 3 TeV            & 400GeV & \multicolumn{2}{c|}{100GeV} \\
  \hline
p power, (MW)    & 4                & 4   &  \multicolumn{2}{c|}{4} \\
$1/\tau_\mu$ (Hz)& 32               & 240 &  \multicolumn{2}{c|}{960} \\
$\mu$/bunch      & $2\times10^{12}$ & $2\times10^{12}$ & \multicolumn{2}{c|}{$2\times10^{12}$} \\
%$\mu$ power (MW) & 28               & 4   & \multicolumn{2}{c|}{1} \\
circumference (m)& 6000             & 1000& \multicolumn{2}{c|}{350} \\
$<B>$ (T)          & 5.2              & 4.7 & \multicolumn{2}{c|}{3} \\
$n^{effective}_{turns}$&785         & 700 & \multicolumn{2}{c|}{450} \\
$\delta p/p$ (\%)& 0.16             & 0.14& 0.12 & 0.003 \\
6-D $\epsilon_{6,N}$
  ($\pi$m$^3$)   &$1.7\times10^{-10}$&$1.7\times10^{-10}$&$1.7\times10^{-10}$&$1.7\times10^{-10}$\\
RMS $\epsilon_T$ ($\pi$ mm-rad)
                 & 0.05            & 0.05 & 0.085& 0.29 \\
$\beta^*$ (cm)   & 0.3             & 2.6  & 4.1  & 14.1 \\
$\sigma_z$ (cm)  & 0.3             & 2.6  & 4.1  & 14.1 \\
$\sigma_r$ ($\mu$m)& 3.2           & 26   & 86   & 294 \\
Luminosity, ($cm^2s^{-1}$)& $7\times10^{34}$ & $10^{33}$ & $1.2\times10^{32}$ & $10^{31}$  \\

  \hline
\end{tabular}
\end{center}
\end{table}

The collider rings radius should be minimized to increase the number
of turns the muons make before they decay. 
For average dipole fields of 5 Tesla, 750 effective turns
are made.
A 115~GeV collider could have a circumference of around 350~m, and 
a 3~TeV machine would be only 6~km in circumference.

However the dipoles do have to cope with the electrons coming from
the muon decay, which, due to their lower momentum are bent onto the
inner wall of the collider. It may be that a substantially open design,
allowing the electrons to emerge from the superconducting magnet
region is optimal, but in any case some cm of shielding will be required.
This is why a 5~Tesla field is considered, rather than the higher
values achieved for LHC tests.

One of the great advantages of the muon collider is the energy precision
and calibration. Due to the reduction of bremsstrahlung a bunch with a
very small energy spread can be maintained, and the two 100~GeV machines
in table~\ref{ta:collider} differ only in this spread. This is required
if a narrow resonance like the Standard Model Higgs around 115~GeV is
to be scanned.

The energy (and its spread) can be measured very accurately,
using the muons spin precession. This is the same measurement as formed
the basis of the LEP calibration, but is much easier because the
muon decay to electron is self analyzing, and allows the measurement of
the polarization on every turn, and this means that each fill can be
calibrated with a precision given by g-2. Also the energy spread can be 
extracted from the dilution of the polarization with time.

\subsection{Detector}

The detector suffers from one major difficulty - the background from 
electrons coming from the muon decay. These will spill into the active
volume creating fake tracks and noise hits, and at high energy Bethe-Heitler
muons will also be created, which are too difficult to stop.
Suppression of the decay electrons will rely upon a complex masking scheme,
and current designs reduce the noise levels to about the same as in LHC 
detectors. Unfortunately this does include a mask in the low angle region,
probably up to 20$^o$ to the axis, which will reduce the physics performance
in the forward region. 
This leads to similar solutions being suggested: pixel detectors
starting at a few cm radius appear to be able to cope with the noise.

The calorimeter may suffer from the Bethe-Heitler muons, and it seems
that a segmented design, where these can be recognized by their orientation,
will be required.

\section{Higgs Physics}

%\subsection{Scan of the Standard Model Higgs}

\begin{figure}[ht]
\center
\unitlength1cm
\begin{picture}(15.5,7.5)
\put(-0.5,1){\epsfig{file=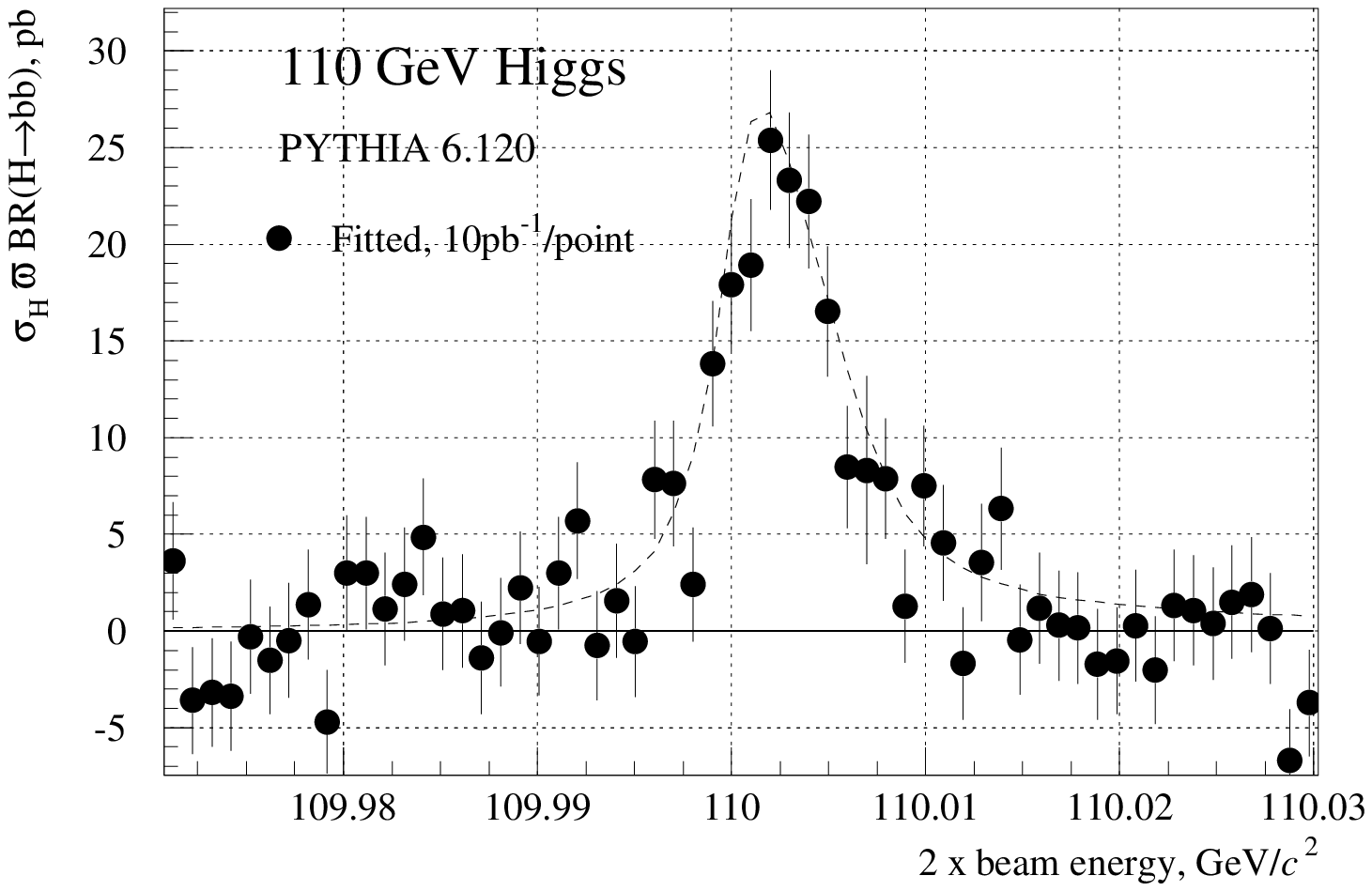,width=0.50\textwidth}}
\put(7.3,0){\epsfig{file=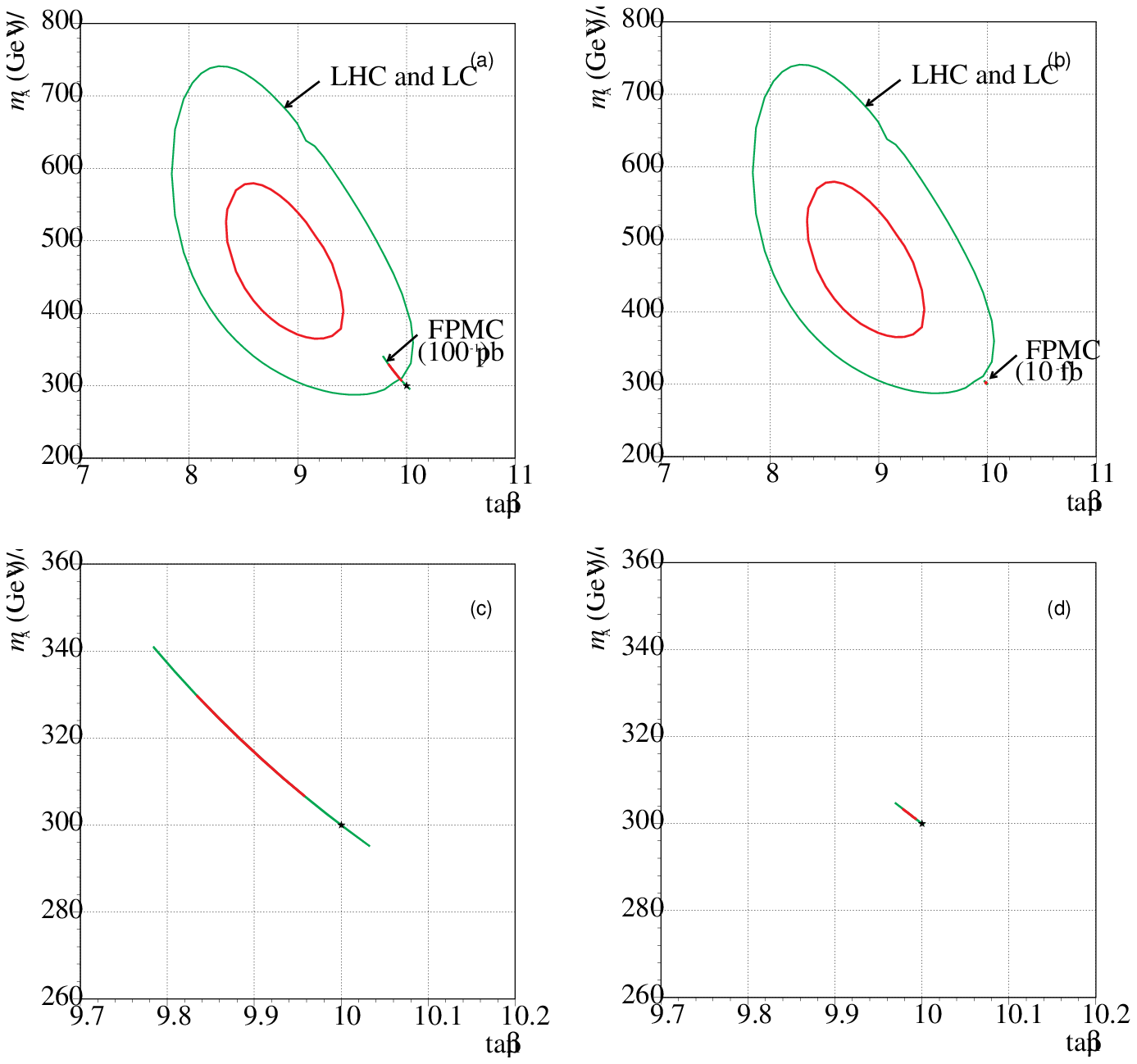,width=0.53\textwidth}}
\put(0.5,6.5){\mbox{\bf a)}}
\put(7.3,6.5){\mbox{\bf b)}}
\end{picture}
\caption{a) The Higgs line-shape, with points corresponding to 10~pb$^{-1}$
 superimposed.
b) The implications of the $h$ scan for the MSSM parameter space, assuming 
all other SUSY parameters are known.
\label{fig:higgs-scan}}
\end{figure}

A sample scan of the Higgs resonance is shown in
figure~\ref{fig:higgs-scan}~a).
The Higgs needs to be found before the collider can be built, and its mass
known. However, if the mass information is accurate to only
60~MeV, as
seems likely\cite{ref:tesla-tdr}, then it will take a year to scan this region and locate 
the resonance. After a couple more years the width will be known to 
1~MeV and the mass error is at present limited by g-2 to about 100~KeV.

The Higgs width is a very interesting test of the model, as it is uniquely
fixed (given the mass) in the Standard Model, but differs in extensions
such as supersymmetry. If the LHC and a linear collider information can
exclude the A of the MSSM (for $\tan\beta > 3$) below 400~GeV, one year
with a muon collider will extend this to 900~GeV, and further with
more luminosity.

%\begin{figure}
%\center
%\label{fig:mssm_parameters}}
%\end{figure}

If the $h$ scan does show something more consistent with the MSSM than 
the SM we can immediately constrain the model parameters.
Figure~\ref{fig:higgs-scan}~b) shows the improvement that the muon collider
would make compared with the information likely to be available from the
LHC and Linear Collider\cite{ref:tesla-tdr}. The plots on the left give the improvement from
a very modest 100~pb$^{-1}$, while those on the right show what could be
learnt in 10 fb$^{-1}$ were available. No theoretical errors have been
allowed for; these are currently substantial\cite{ref:moriond2001-weiglen}.

\subsection{Scan of H and A in the MSSM}

If the standard Model Higgs weighs more than twice the W mass then 
its increased width due to decay to W pairs means that the peak
cross-section is too small for a direct scan to make sense. This is
not however true for the heavier Higgses of supersymmetry, and it 
will be desirable to make  a direct scan of these resonances as well.

\begin{figure}[ht]
\center
\epsfig{file=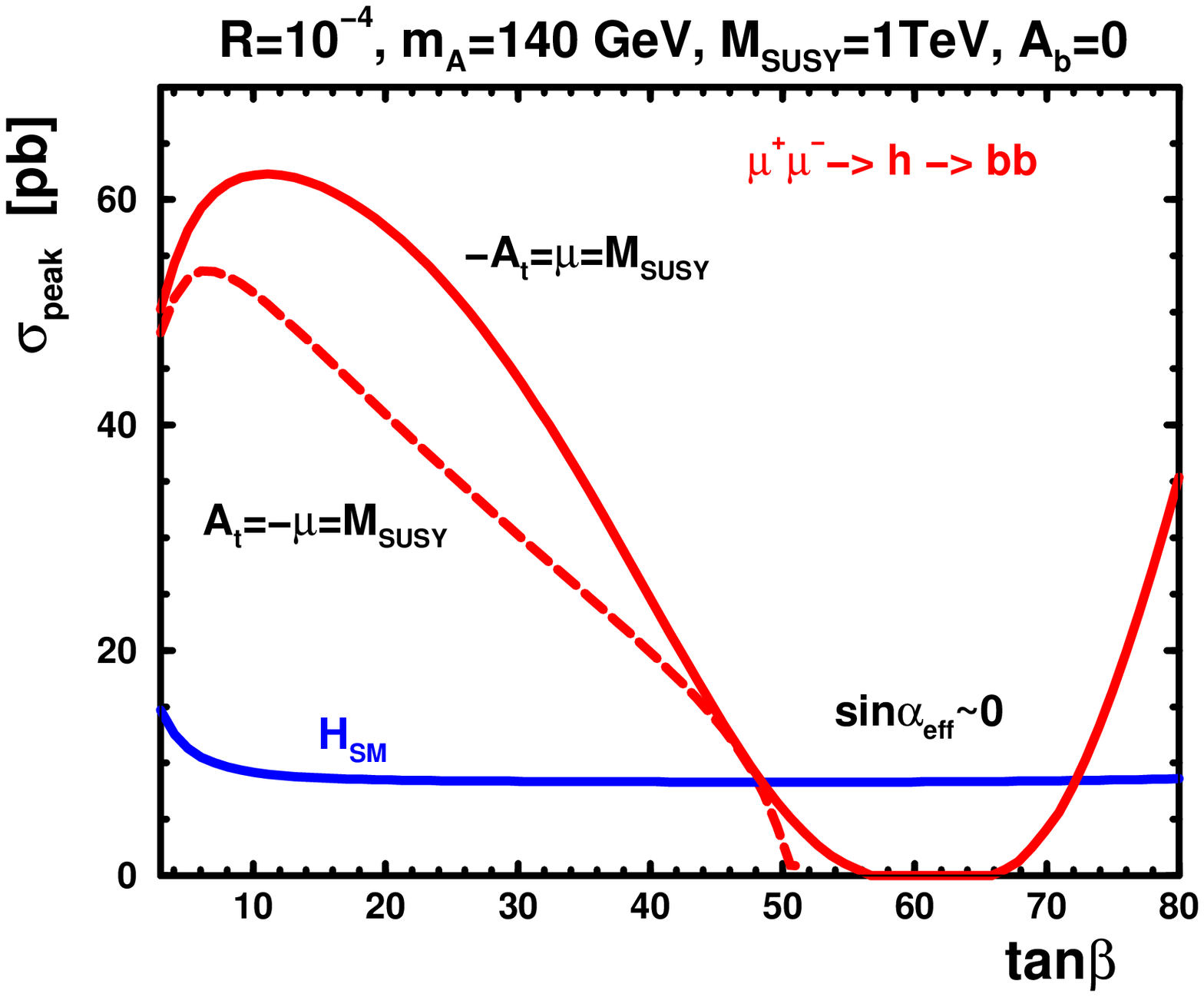,width=0.4\textwidth}
\epsfig{file=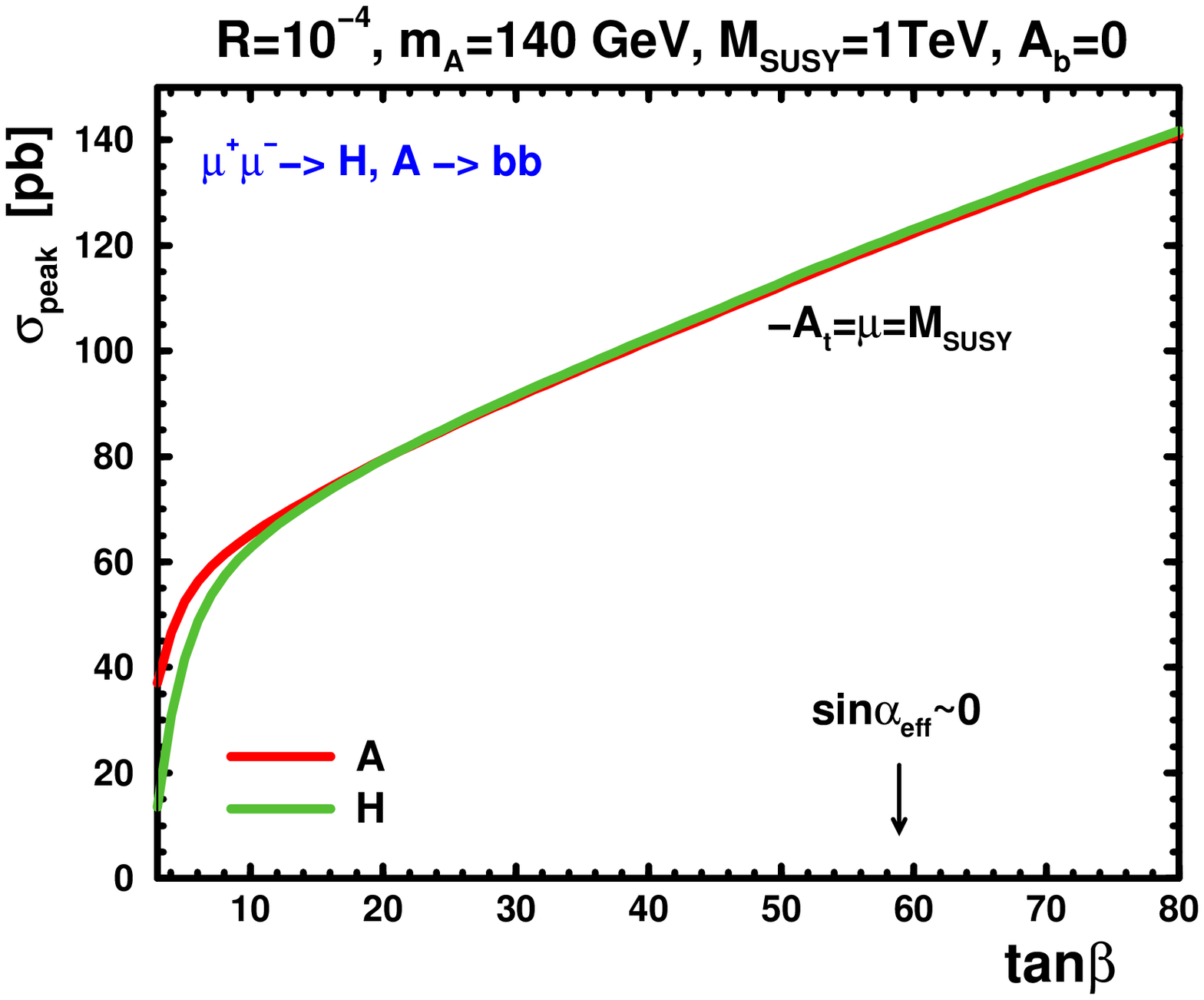,width=0.4\textwidth}
\caption{The dependence of the Higgs couplings on $\tan\beta$, for
a particular choice of MSSM parameters. Note that the SM Higgs mass is
set equal to $m_h$ of the MSSM, and therefore depends upon  $\tan\beta$.
\label{fig:nolose}}
\end{figure}

The coupling of the muon to the different Higgses depends upon the
Higgs mixing angles, for example $g_{h\mu\mu}$ is proportional to
$\sin\alpha_{eff}$, but figure~\ref{fig:nolose}\cite{ref:cern01} makes it
clear that if the lightest is suppressed the heavier Higgses will
be enhanced. Thus the MSSM presents an even more interesting 
picture for a muon collider.

The mass splitting of the H and A is rather small, and it is unlikely
that any other machine can resolve them as separate resonances.
What value it is will depend upon the other parameters, but it
is clear from figure~\ref{fig:nolose} that the cross-sections can 
be similar to or larger than the $h$, and if the widths are relatively
large the collider luminosity can be increased.

\subsection{CP violation}

One very interesting area for study is the CP properties of the
Higgs system. In the MSSM, for example, the h, H and A can in general be 
mixed, and so it is important to measure this. 
If the second Higgs doublet is not so heavy that it decouples
 then the lightest physical Higgs state may well have mixed CP. In this
case there will be very interesting studies,
because production of say $b$ quark pairs can proceed through $\gamma$, Z or
 h exchange, and so interference   can give rise to observable effects.
The heavier states are in general more sensitive, because their masses
are similar and mixing is more likely. 

\begin{figure}[ht]
\center
\epsfig{file=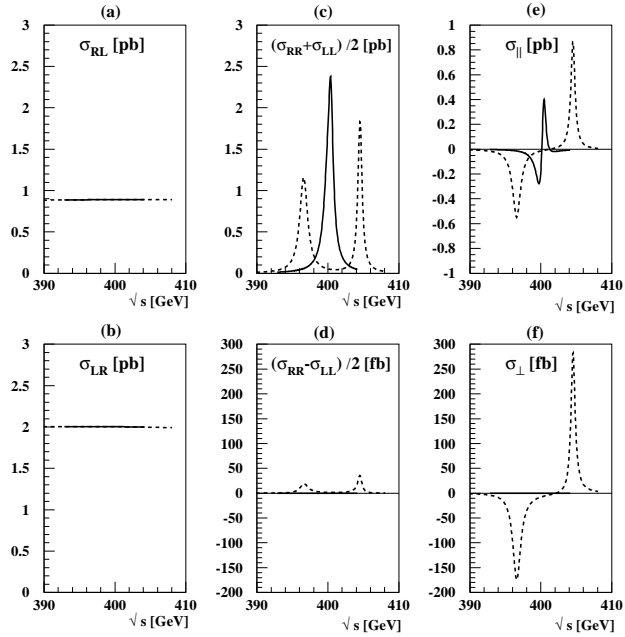,width=0.59\textwidth}
\vspace*{-0.5cm}
\caption{Asymmetries sensitive to CP violation which can be constructed 
in the H, A system.
\label{fig:choi1a}}
\end{figure}

Figure~\ref{fig:choi1a} shows one
analysis\cite{ref:SYC} based on direct CP violation.
Note that for these parameters the introduction of CP violation generates
a mass splitting of the H and A which was not present without.

\section{Summary}

The muon collider operated as a Higgs factory will give the
definitive measurement of the Higgs mass and width. There 
are also unequaled opportunities for establish the CP nature of the
system.
There is a lot of
work to be done before a machine can be constructed, but no insurmountable
obstacles have been identified.
In the meantime the neutrino factory developments will push forward the
technology.

\end{document}